\documentclass{aa}
\usepackage{graphicx}
\usepackage[varg]{txfonts}
\usepackage{pdflscape}
\usepackage{amsmath}
\usepackage{empheq}
\usepackage{nccmath}
\usepackage{gensymb}
\usepackage{amssymb}
\usepackage{textcomp}
\usepackage{xfrac}
\usepackage{xcolor}
\usepackage[colorlinks=true, citecolor=blue, linkcolor=blue, urlcolor=blue]{hyperref}
\usepackage{natbib} 

\defcitealias{Daugevicius2024}{Paper~I}

\begin{document} 
\title{Deriving physical parameters of unresolved star clusters}
\subtitle{IX. Sky background effects in the aperture photometry}
\author{Karolis Daugevi{\v c}ius \and Rima Stonkut{\.e} \and Eimantas Kri{\v s}{\v c}i{\= u}nas \and Erikas Cic{\.e}nas \and Vladas Vansevi{\v c}ius}
\institute{Center for Physical Sciences and Technology, Saul\.{e}tekio av. 3, 10257 Vilnius, Lithuania \\\email{vladas.vansevicius@ftmc.lt}}

\date{Received 14 April 2025; accepted 4 June 2025}

\abstract
{The aperture photometry method is a powerful tool that enables us to study large star cluster systems efficiently. However, its accuracy depends on various factors, including the stochasticity of the stellar initial mass function and variations in the sky background. Previously, in the eighth paper of this series, we established the best achievable limits of the aperture photometry method for star cluster studies in the local universe.}
{The aim of this study is to determine how the sky background affects the limits and applicability of the aperture photometry method in star cluster analysis.}
{We used a large sample of star cluster models spanning the parameter space of \object{M\,31} clusters. Synthetic images were generated to match the Panchromatic $Hubble$ Andromeda Treasury (PHAT) survey observations. To determine how the background affects star cluster photometry, we placed images of simulated clusters into five background fields of different stellar density from the PHAT survey and measured them using aperture photometry.}
{We determined age and mass limits for the \object{M\,31} disc star clusters at which photometric uncertainties are low enough to enable the determination of cluster parameters using the aperture photometry method. We demonstrated that for typical-size clusters, optimal aperture diameters are of $\sim$3 half-light radii. We assessed cluster detection completeness in relation to varying sky background densities, based on the \object{M\,31} PHAT survey data. Our results suggest that a significant selection bias towards more compact clusters may exist in the PHAT survey. We derived low-mass limits of the cluster mass function (CMF) in the PHAT survey, reaching down to masses of $\sim$500~${M_\sun}$ in outer disc areas, $\sim$1500~${M_\sun}$ in middle disc or star-forming regions, and $\sim$3000~${M_\sun}$ in inner disc regions. Therefore, we stress a necessity of careful accounting for selection effects arising due to sky background variations when studying the CMF.}
{}


\maketitle

\section{Introduction}
   \label{Sec1}
Based on current knowledge of how star formation works, the leading belief is that the majority of stars are born in star clusters and associations \citep{Lada2003}, most of which then dissolve into the general stellar population of their host galaxy due to tidal and dynamical perturbations \citep{PortegiesZwart2010}. This makes studies of star clusters ideal for understanding the formation and evolution of galaxies, as well as for constraining theories of star formation processes.

One of the most commonly applied methods for studying extragalactic star clusters involves fitting stochastic models to integrated cluster magnitudes and colour indices (CIs) \citep{Deveikis2008, Fouesneau2010, Fouesneau2014, deMeulenaer2013, Krumholz2015}. Compared to other widely used colour-magnitude diagram (CMD) fitting techniques \citep{Weisz2015, Johnson2016, Wainer2022, Ceponis2024}, aperture (or integrated) photometry methods are suitable to lower-mass, older, and more distant clusters, making them applicable for larger and more diverse samples. However, the accuracy of this approach strongly depends on the precision of aperture photometry measurements.

Previous studies \citep{Fouesneau2010, Beerman2012, Anders2013, deMeulenaer2013, deMeulenaer2017} have shown that a major limiting factor for the accuracy of integrated photometry methods is their sensitivity to stochastic effects. This is particularly pronounced in young ($<$1~Gyr) low-mass ($\lesssim$3000~${M_\sun}$) star clusters, and it is mostly driven by the stochasticity of the stellar initial mass function (IMF). Such clusters tend to have under-sampled IMFs and exhibit sparsely populated upper main-sequence (MS) and post-main-sequence (PMS) evolutionary stages. \citet{Beerman2012} showed that a single bright star, or just a few PMS stars, can introduce significant stochastic variations in a cluster’s brightness and colours, which in turn severely affects the accuracy of cluster age and mass determination.

Another important factor limiting the accuracy of star cluster aperture photometry, especially in the local universe, is the uncertainty in sky background estimation. Multiple $Hubble$ Space Telescope (HST) studies of star clusters in the Andromeda galaxy (\object{M\,31}) \citep{Krienke2007, Krienke2008, Hodge2009, Hodge2010, Johnson2012} showed that the strongly variable background level of the \object{M\,31} disc, its estimation, and subtraction is one of the primary sources of photometric errors. \citet{Kodaira2004} and \citet{Narbutis2008} also reported that photometric uncertainties are predominantly driven by sky background variations and could hardly be reduced without interactive interventions. Moreover, \citet{deMeulenaer2017} showed that the presence of bright field stars within the photometric aperture significantly affects the measurements, complicating the derivation of cluster parameters and often resulting in age-extinction degeneracies. Thus, to ensure reliable interpretation of results derived from star cluster aperture photometry, it is crucial to quantify the limitations imposed by these effects. 

In our previous study, \citeauthor{Daugevicius2024} (\citeyear{Daugevicius2024}, hereafter \citetalias{Daugevicius2024}), we computed a large sample of star cluster models covering the parameter space representative of real \object{M\,31} clusters and generated their synthetic images mimicking observations from the Panchromatic $Hubble$ Andromeda Treasury (PHAT; \citealt{Dalcanton2012}) survey. We performed aperture photometry of those artificial clusters and determined the best achievable accuracy and applicability limits of the aperture photometry method, imposed by the effects of the stochastic IMF. We showed that clusters with and without PMS stars have significant photometric differences, which at young ages ($\sim$10~Myr) complicate the derivation of cluster parameters based on integrated photometry. Furthermore, we demonstrated that reliable physical cluster parameters can be derived when cluster CIs are measured using apertures with diameters larger than their doubled half-light radii.

In this study, we aim to expand on the results presented in \citetalias{Daugevicius2024} by determining how the sky background affects the accuracy and applicability limits of the aperture photometry method for star cluster studies in the local universe. We employ stochastic star cluster models and their realistic HST images, generated using the algorithm described in \citetalias{Daugevicius2024}. To estimate background effects on aperture photometry limits, we placed simulated cluster images into five PHAT survey background fields of different stellar density and measured them using apertures of various sizes. Furthermore, we examined how selection effects may influence the capabilities of present and future star cluster surveys.

The structure of the paper is as follows: in Sect.\,\ref{Sec2}, we briefly present our star cluster models; in Sect.\,\ref{Sec3}, we show and discuss the selected background fields; in Sect.\,\ref{Sec4}, we describe the aperture photometry procedures used to measure our mock observations; in Sect.\,\ref{Sec5}, we present and discuss the results; in Sect.\,\ref{Sec6} we provide a brief summary and conclusions. 

\section{Star cluster models}
	\label{Sec2}

\begin{table}
	\setlength{\tabcolsep}{4pt}
	\caption{\label{t1} Artificial star cluster parameters.}
	\centering
	\begin{tabular}{ll}
		\hline\hline
		Parameter & Values\\
		\hline
		log$({M/M_\sun})$ & 2.0, 2.5, 3.0, 3.5, 4.0. 4.5. 5.0\\
		log$(t/\rm yr)$ & 6.5, 7.0, 7.5, 8.0, 8.5, 9.0, 9.5, 10.0\\
		$r_{c}$\tablefootmark{a} & 0.05, 0.10, 0.20, 0.40, 0.80\\
		$\gamma$ & 2.2, 2.4, 2.8, 3.5, 7.0\\
		\hline
	\end{tabular}
	\tablefoot{\tablefoottext{a}{Cluster core radius (\arcsec), as defined by \citet{King1962}; the relation of the core radius to the EFF scale factor: ${r_{c}}={r_{0}}(2^{\sfrac{2}{\gamma}}-1)^{\sfrac{1}{2}}$}.
	}
	\label{table:tab1}
\end{table}

In this study, we employed artificial star clusters and their images, synthesised using the algorithm described in \citetalias{Daugevicius2024}, to which we refer for a detailed description of the modelling procedure; a brief summary is provided below.

Initial stellar masses are generated according to \citet{Kroupa2001} IMF using fully stochastic sampling in the range from 0.1~${M_\sun}$ to 100~${M_\sun}$. Individual stars are sampled until the total mass within a sphere of 7.5\arcsec\ radius ($\sim$28~pc at the \object{M\,31} distance) reaches the cluster's target mass. For each individual star, PHAT magnitudes are assigned by interpolating the PARSEC-COLIBRI \citep{Bressan2012, Marigo2017} isochrones\footnote{\url{http://stev.oapd.inaf.it/cgi-bin/cmd}} according to its mass and age. We used isochrones of solar metallicity and no interstellar extinction. To convert absolute magnitudes to apparent magnitudes, the \object{M\,31} distance modulus of $(m-M)_{0}=24.47$ \citep{McConnachie2005} was applied. Then stars are spatially distributed within the cluster by stochastically sampling an empirical Elson-Fall-Freeman \citep[EFF;][]{Elson1987} profile, defined by the following radial density function:
\begin{fleqn}
	\begin{align}
	\label{eq:1} \rho(r)=\rho_{0}\left( 1+\dfrac{r^{2}}{r_{0}^{2}}\right)^{-(\gamma+1)/2},
	\end{align}
\end{fleqn}
where $\rho_{0}$ -- the central cluster density; $r$ -- the deprojected (3D) distance from the cluster's centre; $r_{0}$ -- the scale factor; $\gamma$ -- the power-law slope.

To generate realistic mock observations of simulated star clusters, we used the $\tt TinyTim$\footnote{\url{https://www.stsci.edu/hst/instrumentation/focus-and-pointing/focus/tiny-tim-hst-psf-modeling}} package to model images of individual stars. $\tt TinyTim$ enables the modelling of point spread functions (PSFs) for various HST instrument/filter combinations. This representation of PSFs is appropriate for our study, since integrated cluster photometry is performed using apertures that are much larger than the FWHM of PSFs. Each star was simulated by multiplying its total flux in a given passband by the corresponding $\tt TinyTim$ PSF. Images of clusters were generated within $15\arcsec \times 15\arcsec$ frames ($\sim$$57\times57$~pc at the \object{M\,31} distance). 

Following the algorithm described above, we generated a grid of star cluster models and their corresponding images, covering the parameter space representative of real clusters observed in the \object{M\,31} disc. Compared to \citetalias{Daugevicius2024}, the grid was expanded from 500 to 1400 nodes by including more massive clusters and introducing additional ages. Each grid node consists of 100 independently generated artificial clusters. The parameters of the simulated star clusters are listed in Table\,\ref{table:tab1}.

Additionally, for results presented and discussed in Sect.\,\ref{Sec52}, we used a separate set of artificial star clusters simulated using the same modelling approach. However, instead of a discrete grid, we generated a continuous set of artificial clusters in the $F475W$ passband, containing 20 million clusters. The clusters span the following parameter ranges: log$({M/M_\sun})=2.4-4.1$; log$(t/\rm yr)=6.5-10.1$; $r_{c}=0.05\arcsec - 0.6\arcsec$; $\gamma = 2.1-4.1$ \citep{Sableviciute2006, Sableviciute2007}.

\section{Background fields}
	\label{Sec3}

\begin{figure}
	\centering
	\includegraphics[width=9cm]{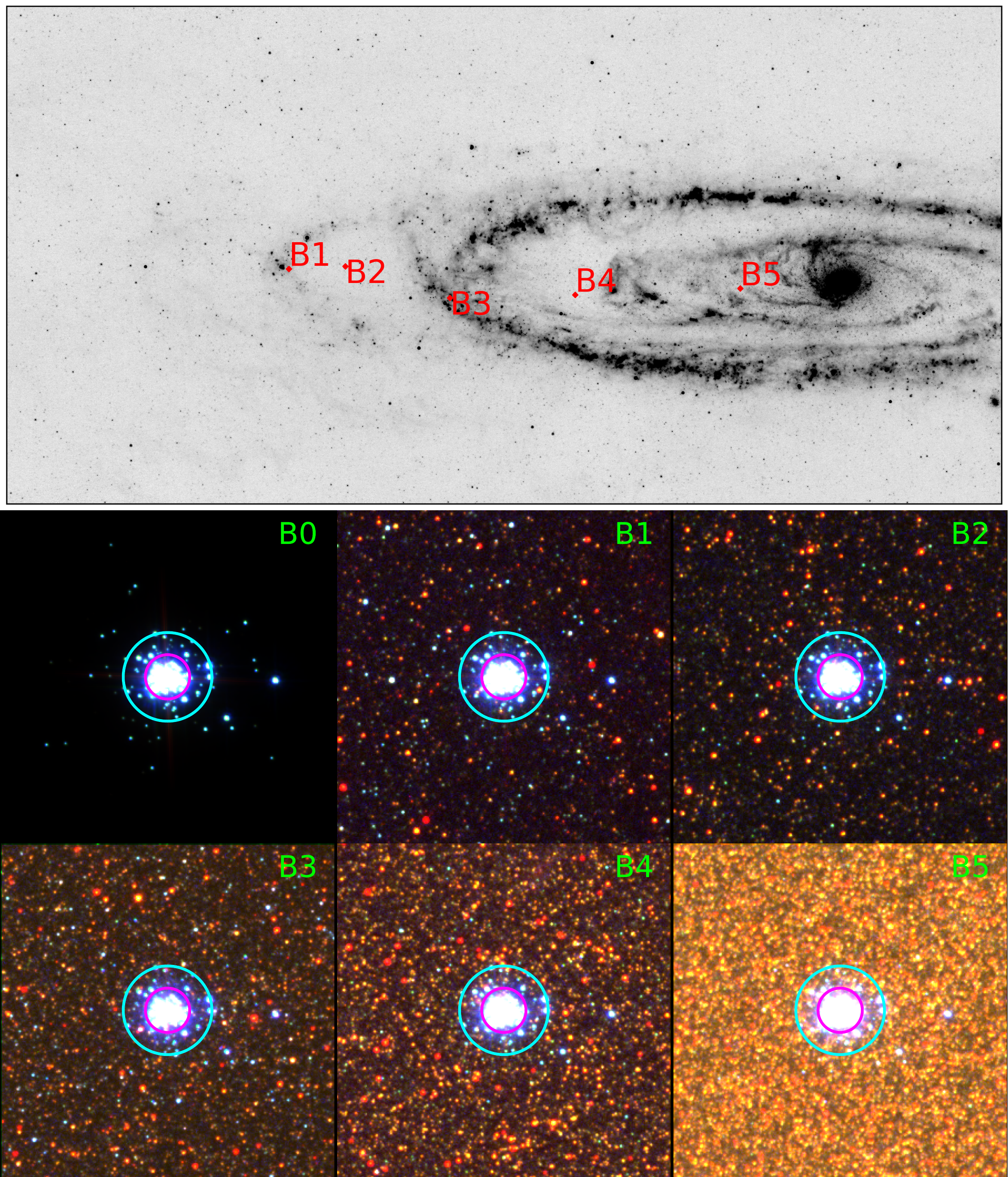}
	\caption{Background fields in \object{M\,31}. Top panel shows the greyscale 24~\textmu m map of \object{M\,31} (Multi-Band Imaging Photometer for \textit{Spitzer}, \textit{Spitzer}/MIPS) with the marked locations of background fields (B1--B5). The bottom panels show images of the artificial star cluster (log$({M/M_\sun})=3.5$, log$(t/\rm yr)=7.0$, $r_{c}=0.2\arcsec$, and $\gamma=2.8$) placed in each of the background fields. B0 -- the background-less case. Magenta and cyan circles mark apertures with radii $R_{\rm ap}=1$\arcsec\ and 2\arcsec, respectively. The colour images ($15\arcsec \times 15\arcsec$) were produced by combining $F336W$, $F475W$, and $F814W$ data.}
	\label{fig1}
\end{figure}

\begin{figure}
	\centering
	\includegraphics[width=9cm]{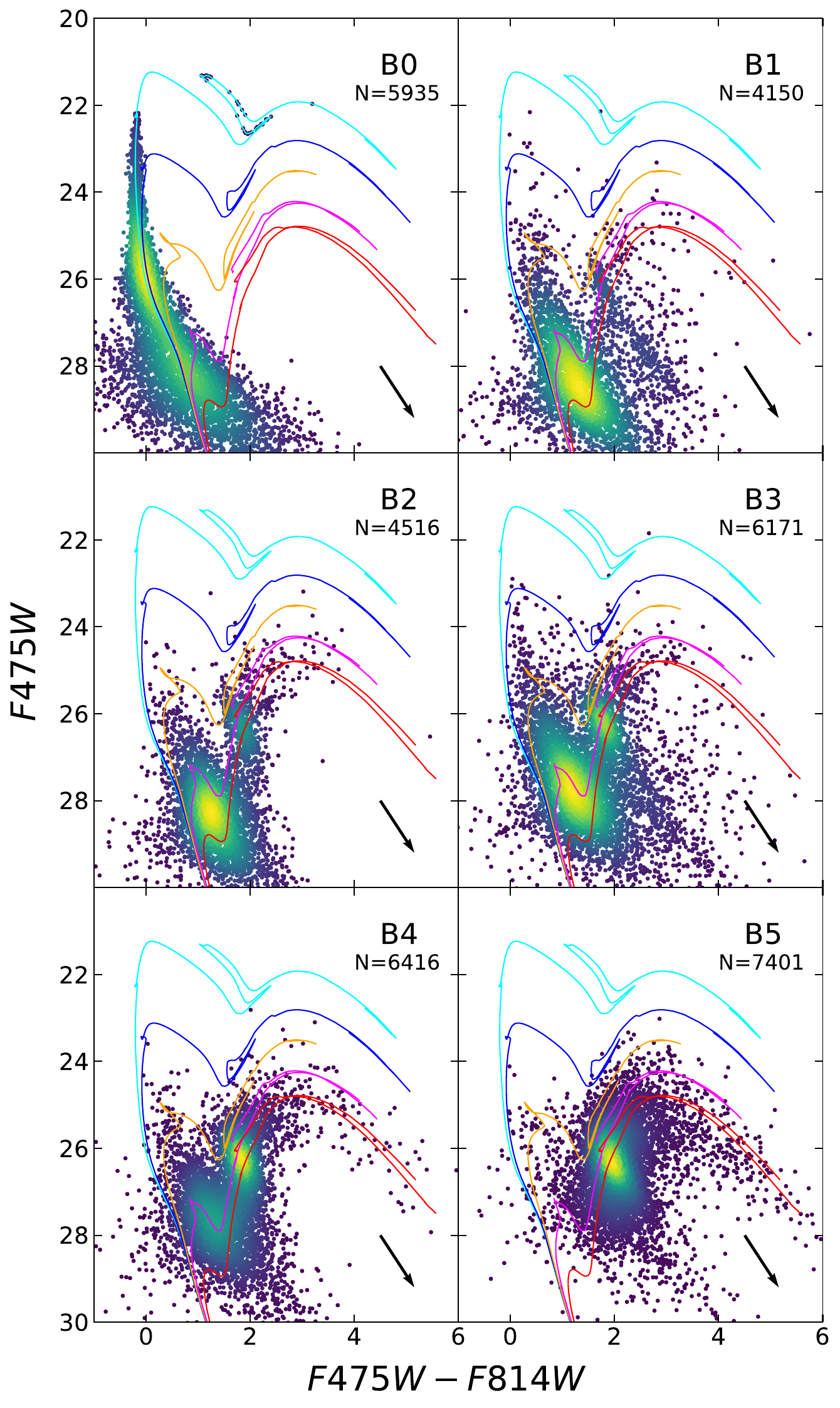}
	\caption{CMDs of the selected background fields. Each panel represents a different field (B0--B5). For the background-less field (B0), the CMD of the generated cluster (log$({M/M_\sun})=4.5$, log$(t/\rm yr)=8.0$), assuming photometry errors from \citet{Williams2023}, is shown. Stellar photometry for fields B1--B5 is taken from \citet{Williams2023}. The number of stars visible in the CMD is indicated below the field name. Colour curves represent the PARSEC-COLIBRI solar metallicity isochrones for log$(t/\rm yr)=8.0$ (cyan), 8.5 (blue), 9.0 (orange), 9.5 (magenta), and 10.0 (red) ages. The black arrows mark the extinction vectors, calculated using the \citet{Fitzpatrick1999} extinction law with $A_{V}=1$.}
	\label{fig2}
\end{figure}

To assess the impact of background effects on aperture photometry of star clusters and on catalogue completeness limits in the context of the PHAT survey, we placed artificial clusters into five real background fields of different stellar density in the \object{M\,31} disc (see Fig.\,\ref{fig1}). We used the same PHAT survey image data as in the studies by \citet{Naujalis2021} and \citet{Krisciunas2023}. The full dataset consists of six broad-band filters from three HST instruments, covering a spectral range from 0.25 to 1.6~\textmu m: the $F275W$ and $F336W$ passbands from the Wide Field Camera 3 UVIS  channel (WFC3/UVIS); the $F475W$ and $F814W$ passbands from the Wide Field Channel of the Advanced Camera for Surveys (ACS/WFC); and the $F110W$ and $F160W$ passbands from the Wide Field Camera 3 IR channel (WFC3/IR). For information on the data processing, refer to \citet{Naujalis2021}.

Because only two repeated exposures are available for the $F275W$ and $F336W$ passbands,  these frames are contaminated with numerous cosmic-ray artefacts. Since reliable automatic cleaning is challenging in such cases, we manually removed the most obvious and bright cosmic-ray defects from the selected background fields using the $\tt imedit$ task from $\tt PyRAF$\footnote{\url{https://iraf-community.github.io/pyraf.html}}.

For the background fields (Fig.\,\ref{fig1}), we used square cut-outs of $17\arcsec \times 17\arcsec$ ($\sim$$65\times65$~pc at the \object{M\,31} distance). As these cut-outs are larger than the cluster images, we were able to place each cluster at a slightly different random location within a given background field, while preserving the same overall background features. This approach also helped mitigate potential biases that could be introduced by placing all artificial clusters in the same position, which might contain some peculiar features such as a very bright background star or a dusty patch.

The selected background fields differ not only in stellar density, which mostly depends on the radial distance from the galactic centre, but also in the type of environment they represent. As shown in the \textit{Spitzer} map (Fig.\,\ref{fig1}), two fields (B1 and B3) are located in the star-forming rings, while B2 and B4 lie in the inter-ring regions; meanwhile, B5 represents the dense inner regions of the \object{M\,31} disc. Fig.\,\ref{fig2} shows CMDs of these fields. Due to their location in star-forming regions, fields B1 and B3 have prominent young MS populations, while the remaining fields are dominated by old stellar populations, such as red giant branch (RGB), asymptotic giant branch (AGB), and fainter MS stars. The dependence of the limiting magnitude on the stellar density is clearly visible. In the B5 field, a prominent main sequence is not visible, as the MS turn-off point lies below the limiting magnitude due to the crowdedness of old stellar populations. In addition, all background fields show some level of extinction, as evidenced by observed MSs being shifted to the red relative to the isochrones' main sequences in the B1--B5 panels of Fig.\,\ref{fig2}. Another important aspect is the strong differential extinction effect in fields B1 and B3, which becomes particularly evident when comparing the red clumps of B1 with B2, and B3 with B4. Fields B1 and B3 exhibit strongly dispersed red clump populations, spanning multiple magnitudes in both brightness and colour. The general direction of this dispersion is similar to that of the extinction vector. This, combined with the fact that both fields are located in the star-forming rings, suggests that they are affected by differential extinction caused by dusty lanes and patches, which are prominent in star-forming regions. As the majority of young clusters reside in such regions, it is likely that at least some are affected by differential extinction. It would distort integrated photometric measurements and severely complicate the derivation of cluster parameters \citep{Narbutis2007a, Bridzius2008}. Particularly striking is the fact that such strong differential extinction effects are observed within a relatively small field of view.

The B3 field is located within the 10~kpc star-forming ring, where the majority of clusters from the PHAT survey are found. For this reason, in Sect.\,\ref{Sec5}, we use it (extensively) as the representative field when investigating various parameter dependencies.

\section{Aperture photometry}
	\label{Sec4}
	
In this section, we describe the aperture photometry procedures used in this study. As the first step, we determined the sky background level to be subtracted from the flux measured within a given aperture. We followed the background determination procedure described by \citet{Johnson2012}, which we consider appropriate for this work, as our simulated observations mimic those of the PHAT survey. This method was also used for producing original photometric star cluster catalogues of the PHAT and PHATTER surveys \citep{Johnson2012, Johnson2015, Johnson2022}. A brief overview of the procedure is provided below.

Once an artificial star cluster is placed into a background field, a circular ring extending from $1.2\times R_{\rm ap}$ to $3.4\times R_{\rm ap}$ \citep{Johnson2012} is defined and positioned at the centre of the cluster. In all cases, for sky background level determination we assume $R_{\rm ap}=2.0$\arcsec. That large ring is then divided into 10 smaller annuli of equal area. Each of these has the same area as a circular aperture with $R_{\rm ap}=2.0$\arcsec.

Then we measure the integrated flux within each of the 10 annuli. An iterative 2$\sigma$-clipping procedure is subsequently applied to these flux values. This helps to reject regions affected by bright background stars (or, in some cases, even artificial cluster bright stars, since it is already inserted into the field) or other peculiar features. As the final sky background value, we use the mean of the non-rejected fluxes, while the standard deviation of these values is adopted as the background level determination uncertainty, $\sigma_{s}$. Since we measure artificial clusters using not only an aperture of radius $R_{\rm ap}=2.0$\arcsec, but also various smaller and larger apertures, we scale the mean sky background estimate accordingly to match the corresponding aperture size. We emphasize that the sky background level is determined individually for each artificial cluster's frame after the simulated cluster is already placed into the background field. We consider this approach to be less idealised and more representative of real observational conditions, as the background level estimation can also be affected by cluster stars located in its outer regions.

Once the sky background level is determined, we perform aperture photometry of the artificial cluster placed in a given background field using multiple apertures of different sizes, ranging from $R_{\rm ap}=0.5$\arcsec\ to 5.0\arcsec. Such procedures are performed on every generated star cluster, in all five background fields, with measurements carried out in three PHAT passbands -- $F336W$, $F475W$ and $F814W$. Using the same apertures, we also perform measurements of these clusters in the background-less field (B0) as a reference for quantifying background effects. All aperture photometry procedures, including the sky background level estimation, were performed using tools provided in the $\tt Python$ package $\tt photutils$\footnote{\url{https://doi.org/10.5281/zenodo.596036}}.

We note that in Sect.\,\ref{Sec5} we mostly use the cluster's half-light radius ($R_{50}$) units to define the aperture size. For each individual artificial cluster, $R_{50}$ was determined by measuring its growth curve in the $F475W$ passband up to 5\arcsec\ and evaluating the radius at which 50\% of the total integrated flux inside $R_{\rm ap}=5.0$\arcsec\ aperture is reached. We emphasize that when $R_{50}$ is used to define aperture size and the resulting aperture radius is smaller than 0.5\arcsec, we ensure it is at least 0.5\arcsec. For example, if a cluster has $R_{50}=0.3$\arcsec\ and the chosen aperture is $R_{\rm ap}=1.5 \times R_{50}$, the resulting aperture radius would be $R_{\rm ap}=0.45$\arcsec. However, in such cases, we increase it up to $R_{\rm ap}=0.5$\arcsec. This is justified in \citetalias{Daugevicius2024}, where we showed that colour indices are strongly affected by stochastic effects if small apertures ($R_{\rm ap} < 0.5$\arcsec) are used. Additionally, we have demonstrated that such small apertures lead to unreliably determined physical star cluster parameters.

\section{Results and discussion}
	\label{Sec5}

\subsection{Photometric uncertainties}
\label{Sec51}

\begin{figure*}
	\centering
	\includegraphics[width=18.0cm]{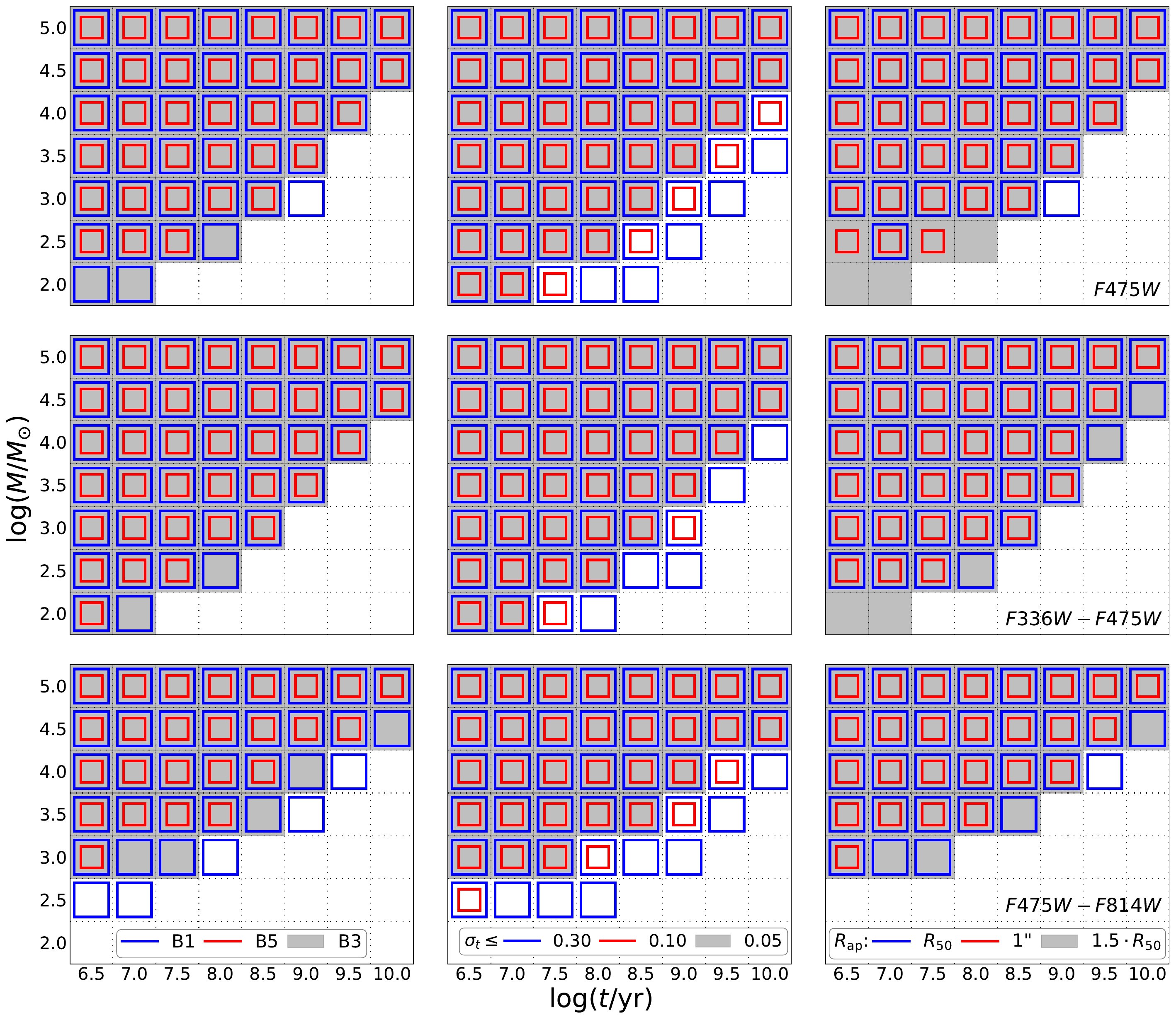}
	\caption{Regions of star cluster parameter space where aperture photometry meets the required accuracy. The grey areas and regions with blue or red squares indicate the cluster mass and age combinations containing at least 50\% (out of $\sim$1300) of clusters satisfying the photometric error threshold, $\sigma_{t}$. Top row -- results for the $F475W$ passband; middle row -- results for $F336W-F475W$; bottom row -- results for $F475W-F814W$. Left column -- results for different background fields, using $R_{\rm ap}=1.5 \times R_{50}$ and the error threshold $\sigma_{t}\leq0.05$~mag. Middle column -- results depending on the $\sigma_{t}$ threshold value for the B3 field, using $R_{\rm ap}=1.5 \times R_{50}$. Right column -- results depending on the aperture size for the B3 field and the error threshold $\sigma_{t} \leq 0.05$~mag. Results are based on artificial clusters with half-light radii $R_{50}=0.25\arcsec - 1.00\arcsec$.}
	\label{fig3}
\end{figure*}

\begin{figure}
	\centering
	\includegraphics[width=9.0cm]{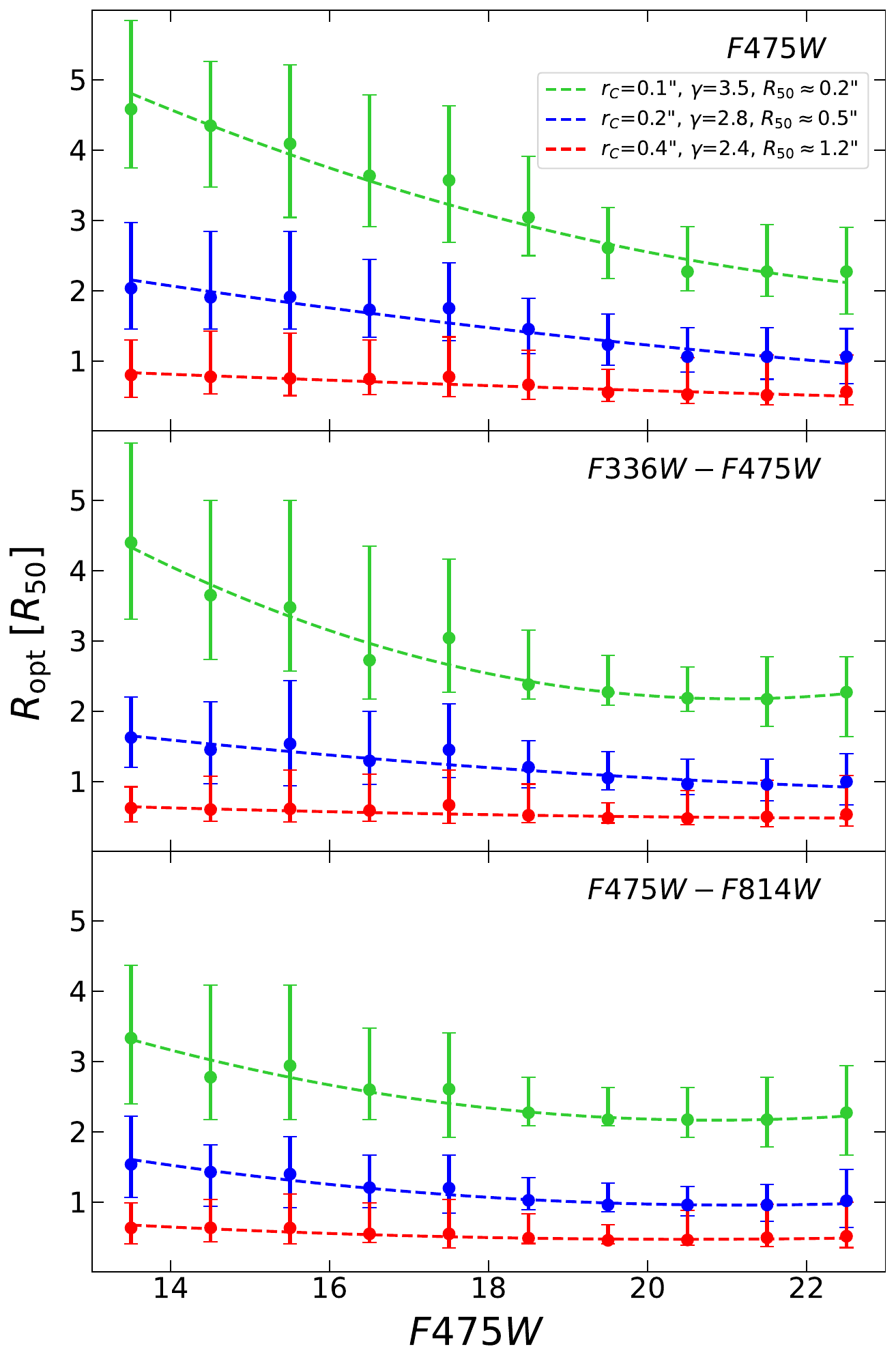}
	\caption{Optimal aperture radii $R_{\rm opt}$ versus the $F475W$ cluster magnitude measured in the B0 field. Top panel: $R_{\rm opt}$ for measuring the $F475W$ magnitude; middle panel -- $F336W-F475W$; bottom panel -- $F475W-F814W$. Results are shown for the B3 field using three combinations of geometric parameters, $r_{c}$ and $\gamma$ (see legend, where approximate median half-light radii $R_{50}$ are also included for reference). Markers show the $R_{\rm opt}$ median values, error bars indicate the 16th--84th percentile ranges. Dashed lines represent second order polynomial least-squares fits.}
	\label{fig4}
\end{figure}

\begin{figure}
	\centering
	\includegraphics[width=9.0cm]{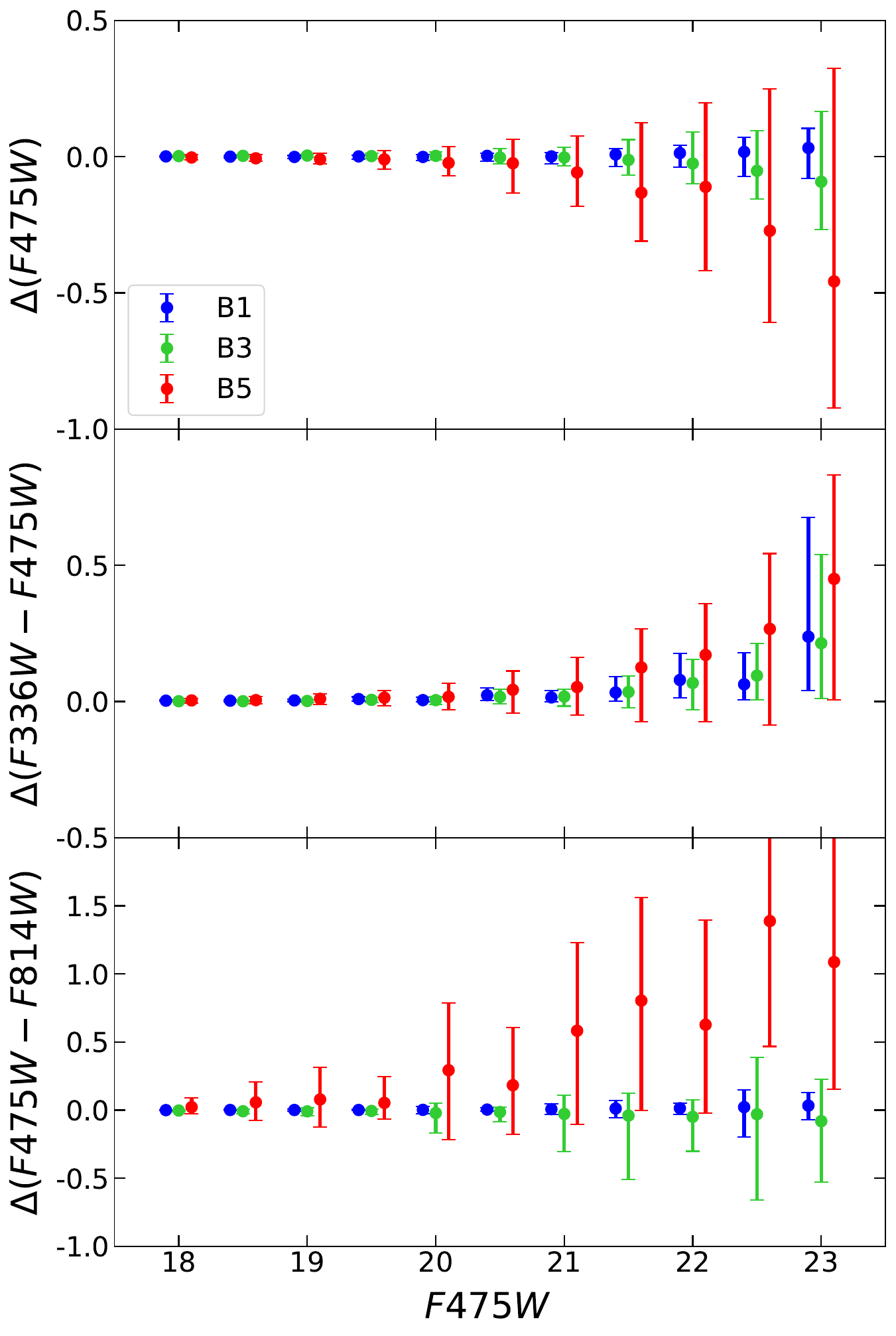}
	\caption{Differences in $F475W$ (top), $F336W-F475W$ (middle), and $F475W-F814W$ (bottom), calculated as measurements in the B1, B3, and B5 fields minus those in the B0 field, versus the $F475W$ cluster magnitude measured in the B0 field. Markers show the median differences; error bars indicate the 16th--84th percentile ranges. Blue and red markers are slightly offset along the $X$-axis for clarity. Results are based on artificial clusters with half-light radii $R_{50}=0.4\arcsec - 0.6\arcsec$.}
	\label{fig5}
\end{figure}

\begin{figure}
	\centering
	\includegraphics[width=9.0cm]{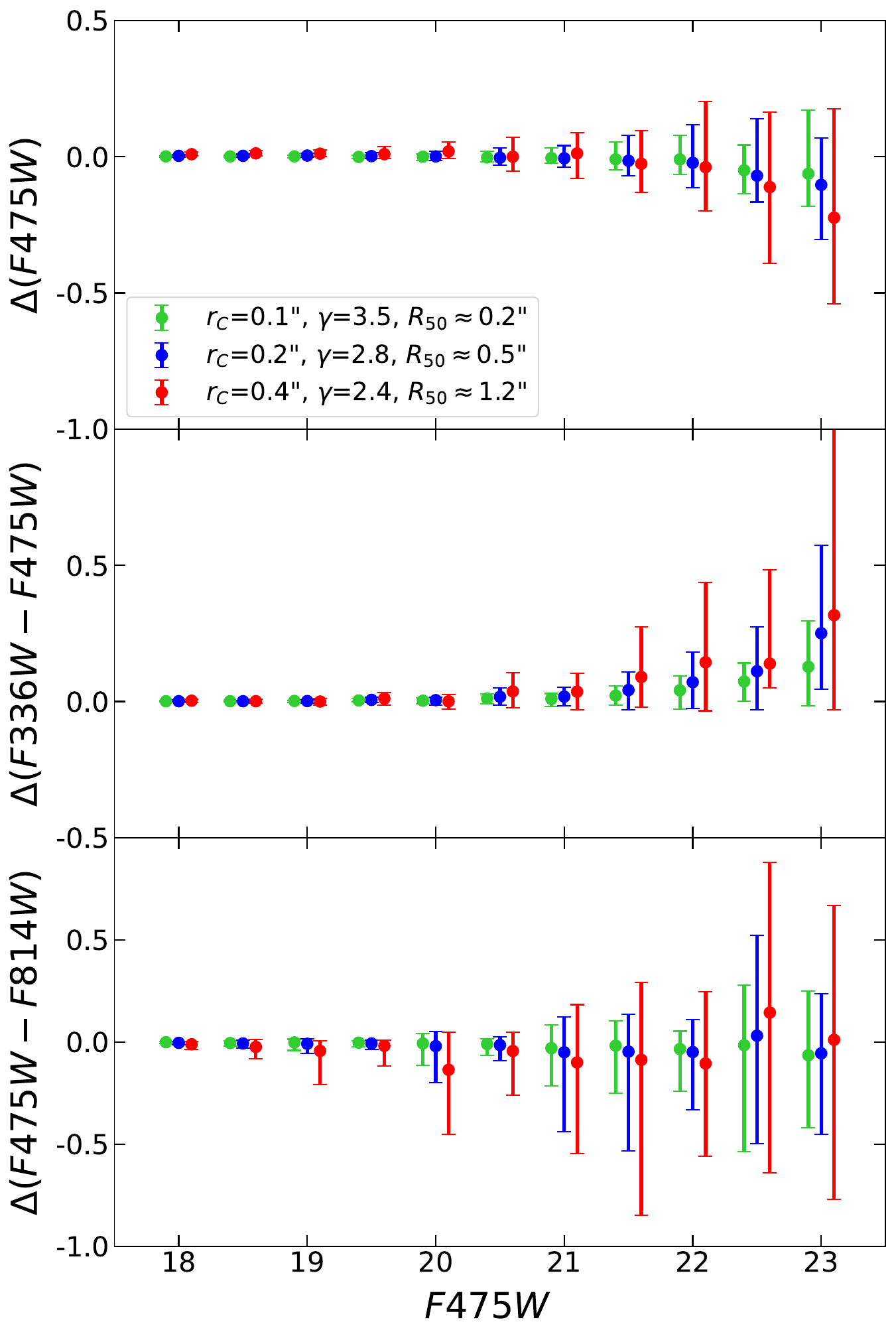}
	\caption{Differences in $F475W$ (top), $F336W-F475W$ (middle), and $F475W-F814W$ (bottom), calculated as measurements in the B3 field minus those in the B0 field, versus the $F475W$ cluster magnitude measured in the B0 field. Results are shown for the same combinations of geometric parameters ($r_{c}$ and $\gamma$) as in Fig.\,\ref{fig4} (see legend, where approximate median half-light radii $R_{50}$ are also included for reference). Markers show the median differences; error bars indicate the 16th--84th percentile ranges. Blue and red markers are slightly offset along the $X$-axis for clarity.}
	\label{fig6}
\end{figure}

In this study, we account for three different error sources in star cluster aperture photometry measurements. One of these is Poisson noise. However, since we used drizzled images and fluxes scaled to one-second exposures, it is problematic to determine these statistical errors correctly. Thus, to roughly estimate the impact of this error source, we multiplied one-second exposure fluxes by the PHAT exposure times provided by \citet{Williams2014}. The Poisson uncertainty for the magnitude measurements is evaluated using the following formula:
\begin{fleqn}
	\begin{align}
		\label{eq:2} \sigma_{p} = 2.5 \log_{10} \left( 1 + \frac{\sqrt{S_{cl}}}{S_{cl}} \right),
	\end{align}
\end{fleqn}
where $S_{cl}$ is the cluster signal (one-second flux $F_{cl}$ multiplied by the exposure time) measured through the given aperture, in the background-less field (B0).

The second source of uncertainties we consider is the effect of possible aperture position bias. We estimated it following the approach described in \citet{Naujalis2021} and \citet{Krisciunas2023}. For each simulated image, we performed aperture photometry by positioning the centre of the selected aperture at each of the nine central pixels of the frame (in the pixel scale of HST/ACS). This gave us 9 cluster magnitude and CIs measurements of each mock observation. Magnitude uncertainties due to this possible aperture position bias ($\sigma_{a}$) were estimated as a half of the 16th--84th percentile range. We emphasize that the $\sigma_{a}$ uncertainty for CIs is determined from 9 CI measurements, rather than by combining the uncertainties of the magnitudes in the individual passbands that comprise the CI.

The third source of errors we consider, arises from the background level estimation procedure. It is evaluated as follows:  
\begin{fleqn}
	\begin{align}
		\label{eq:3} \sigma_{b} = 2.5 \log_{10} \left(1 + \frac{\sigma_{s}\cdot A_{\rm ap}}{F_{cl}} \right),
	\end{align}
\end{fleqn}
where $\sigma_{s}$ is the background level determination uncertainty, which was introduced in Sect.\,\ref{Sec4}, $A_{\rm ap}$ -- the area of the selected aperture in pixels, $F_{cl}$ -- the integrated cluster flux within that aperture measured in the B0 field.

The aggregated photometric uncertainty of the magnitude in a given passband was calculated according to the following formula:
\begin{fleqn}
	\begin{align}
		\label{eq:4} \sigma_{t} = \sqrt{\sigma_{p}^2 + \sigma_{b}^2 + \sigma_{a}^2}.
	\end{align}
\end{fleqn}

Meanwhile, the total photometric uncertainties for CIs were determined by adding squared $\sigma_{p}$ and $\sigma_{b}$ of the individual passbands that comprise the given CI, together with squared $\sigma_{a}$ determined for that CI.

Among the individual sources of uncertainty, we find that the background estimation error ($\sigma_{b}$) is the dominant contributor in the majority of cases, which is in agreement with previous HST studies of \object{M\,31} star clusters \citep{Krienke2007, Krienke2008, Hodge2009, Hodge2010, Johnson2012}. However, it is noteworthy that in some specific cases, the aperture centring error ($\sigma_{a}$) can exceed  $\sigma_{b}$. This typically occurs when a bright cluster member or a background star is projected near the edge of the aperture. The uncertainty due to Poisson noise ($\sigma_{p}$) is relatively insignificant compared to the other two uncertainty sources. In the following analysis, we consider only the aggregated total uncertainties for magnitude and for CIs, both of which are referred to as $\sigma_{t}$ in a general manner. 

The distributions of cluster age and mass combinations that include at least 50\% of models meeting indicated requirements (for $F475W$, $F336W-F475W$, and $F475W-F814W$) are shown in Fig.\,\ref{fig3}, as functions of background fields (B1, B3, B5), photometric error ($\sigma_{t}$) thresholds, and aperture sizes ($R_{\rm ap}$). We applied three error thresholds ($\sigma_{t}=0.05, 0.10, 0.30$) following \citet{deMeulenaer2015}, who discussed that integrated photometry with $\leq$0.05~mag accuracy enables the derivation of cluster age, mass, and extinction, and also, allows estimating metallicity for clusters older than 1~Gyr. Correspondingly, in cases of $\leq$0.10~mag photometric accuracy, we can expect to estimate cluster age, mass, and extinction for known metallicity; in cases of $\leq$0.30~mag photometric accuracy, we can expect to roughly estimate cluster age and mass, for known extinction and metallicity.

The dependence on background environment is most clearly seen in the left column of Fig.\,\ref{fig3}, where results from measurements performed using the aperture of $R_{\rm ap}=1.5 \times R_{50}$ size in the fields B1, B3, and B5 are compared, assuming an error threshold of $\sigma_{t}\leq0.05$~mag. Based on the results for $F475W$ and $F336W-F475W$, we can state that limits of cluster age-mass parameter space, where photometric accuracy, required for physical parameter determination, can be achieved, barely depends on the background density in the UV-optical parts of the spectrum. However, significant dependence is observed in the case of $F475W-F814W$. This is due to numerous bright red stars residing in the \object{M\,31} disc, whose number density is highly dependent on the radial distance from the galactic centre.

Therefore, the most limiting factor is our ability to perform precise photometry in the $F814W$ passband due to numerous bright red background/foreground stars. More accurate photometric measurements in (semi-)resolved cases could be achieved using sophisticated background estimation techniques, such as interactive methods \citep{Naujalis2021, Krisciunas2023} or advanced automated algorithms \citep{Pushpak2025}.

The middle column (Fig.\,\ref{fig3}) shows age and mass limits where at least 50\% of models pass the given photometric accuracy thresholds ($\sigma_{t}=0.05, 0.10, 0.30$). The results are provided for the B3 field, using an aperture of $R_{\rm ap}=1.5 \times R_{50}$. Taking into consideration results for both $F336W-F475W$ and $F475W-F814W$ CIs and interpolating linearly between mass limits derived for each of these CIs, we find that at 1~Gyr, photometric accuracy of $\sigma_{t}=0.05$ can be achieved for $\gtrsim$5000~${M_\sun}$ clusters; $\sigma_{t}=0.10$ -- for $1600 \lesssim {M} \lesssim 5000$~${M_\sun}$ clusters; $\sigma_{t}=0.30$ -- for $500 \lesssim {M} \lesssim 1600$~${M_\sun}$ clusters. However, at 100~Myr lower mass limits can be reached: $\sigma_{t}=0.05$ is attainable for $\gtrsim$1000~${M_\sun}$ clusters; $\sigma_{t}=0.10$ -- for $500 \lesssim {M} \lesssim 1000$~${M_\sun}$ clusters; $\sigma_{t}=0.30$ -- for $200 \lesssim {M} \lesssim 500$~${M_\sun}$ clusters.

The rightmost column (Fig.\,\ref{fig3}) shows how the age-mass parameter space that meets the photometric precision requirement $\sigma_{t}\leq0.05$~mag depends on aperture size $R_{\rm ap}$. Here, we compare measurements performed in the B3 field using three different  aperture sizes ($R_{\rm ap}$): $1\arcsec, R_{50}, 1.5 \times R_{50}$. We notice that aperture sizes scaled by an individual cluster's $R_{50}$ yields better results than a fixed one-for-all aperture ($R_{\rm ap}=1$\arcsec\ in this case). We find that the best results are achieved using $R_{\rm ap}=1.5 \times R_{50}$ apertures, which is consistent with our findings in \citetalias{Daugevicius2024}. It is noteworthy, however, that in some cases, $R_{\rm ap}=R_{50}$ performs better than $R_{\rm ap}=1.5 \times R_{50}$. Nevertheless, as shown in \citetalias{Daugevicius2024}, measurements with $R_{\rm ap} < R_{50}$ are susceptible to significant stochastic effects, especially for small clusters ($R_{50} \approx 0.5\arcsec$), and can result in unreliable estimates of cluster physical parameters. Therefore, we recommend that such small apertures be used with caution, and only in certain special cases, for instance, to avoid contamination from bright field stars projecting near the cluster \citep{Narbutis2007b}.

To expand recommendations on appropriate aperture sizes for cluster colour measurements, we determined the optimal aperture radius ($R_{\rm opt}$) for each mock observation -- defined as the aperture that yields the smallest $\sigma_{t}$. In Fig.\,\ref{fig4} we show how $R_{\rm opt}$ (expressed in units of $R_{50}$) depends on cluster luminosity for three cluster geometric parameter ($r_{c}$, $\gamma$) combinations. The green markers represent compact clusters, blue -- typical \object{M\,31} clusters, and red -- extended ones. The optimal aperture, $R_{\rm opt}$, versus $F475W$ within the 10-magnitude interval differs by a factor of approximately 1.5--2.0 between the faintest and the brightest star clusters. The optimal aperture radius for measuring CIs of typical \object{M\,31} disc star clusters is (1--2)$\times R_{50}$, which is consistent with our recommendations given in \citetalias{Daugevicius2024}. 

In Fig.\,\ref{fig5}, we show differences $\Delta(F475W)$, $\Delta(F336W-F475W)$, and $\Delta(F475W-F814W)$), defined as the magnitudes and colour indices of artificial star clusters measured in the B1, B3, and B5 fields minus those obtained for the same clusters in the B0 field. We find that, beyond a certain magnitude limit (which depends on the background field), systematic deviations from the B0 case begin to appear, increasing with decreasing cluster brightness. Also, the measurements of faint clusters placed in the background fields tend to be systematically brighter compared to the magnitudes measured in the B0 field. Similarly, CI measurements are often biased towards redder values. However, a notable exception is observed in the B3 field, where $F475W-F814W$ shows a slight bias towards bluer values. This may be explained by the presence of a significant population of relatively young, blue stars in that field. It is also noteworthy, that background effects for $F475W-F814W$ are stronger and the bias begins to appear at brighter magnitude compared to $F475W$ or $F336W-F475W$. This is due to the brighter \object{M\,31} background in the $F814W$ passband, resulting in a lower contrast between the cluster and the background signal than in $F336W$ or $F475W$. Such biases can lead to inaccuracies in the determination of physical parameters: a biased higher luminosity may result in an overestimated mass, while redder CIs can lead to an overestimated age, metallicity, or extinction.

In Fig.\,\ref{fig6} we show the same analysis as in Fig.\,\ref{fig5}, but for the B3 field and three different cluster sizes. We find that, at a certain cluster magnitude, larger clusters (red dots) exhibit stronger biases and broader magnitude and CI difference spreads compared to more compact ones. Moreover, for larger clusters, systematic deviations begin to appear at brighter magnitudes. This is due to the use of larger apertures for extended clusters, which include more background signal, thereby reducing the cluster-to-background contrast. Additionally, the probability of a bright background/foreground star falling within the aperture increases with aperture size.

\subsection{Completeness and selection effects}
\label{Sec52}

\begin{figure*}
	\centering
	\includegraphics[width=18.5cm]{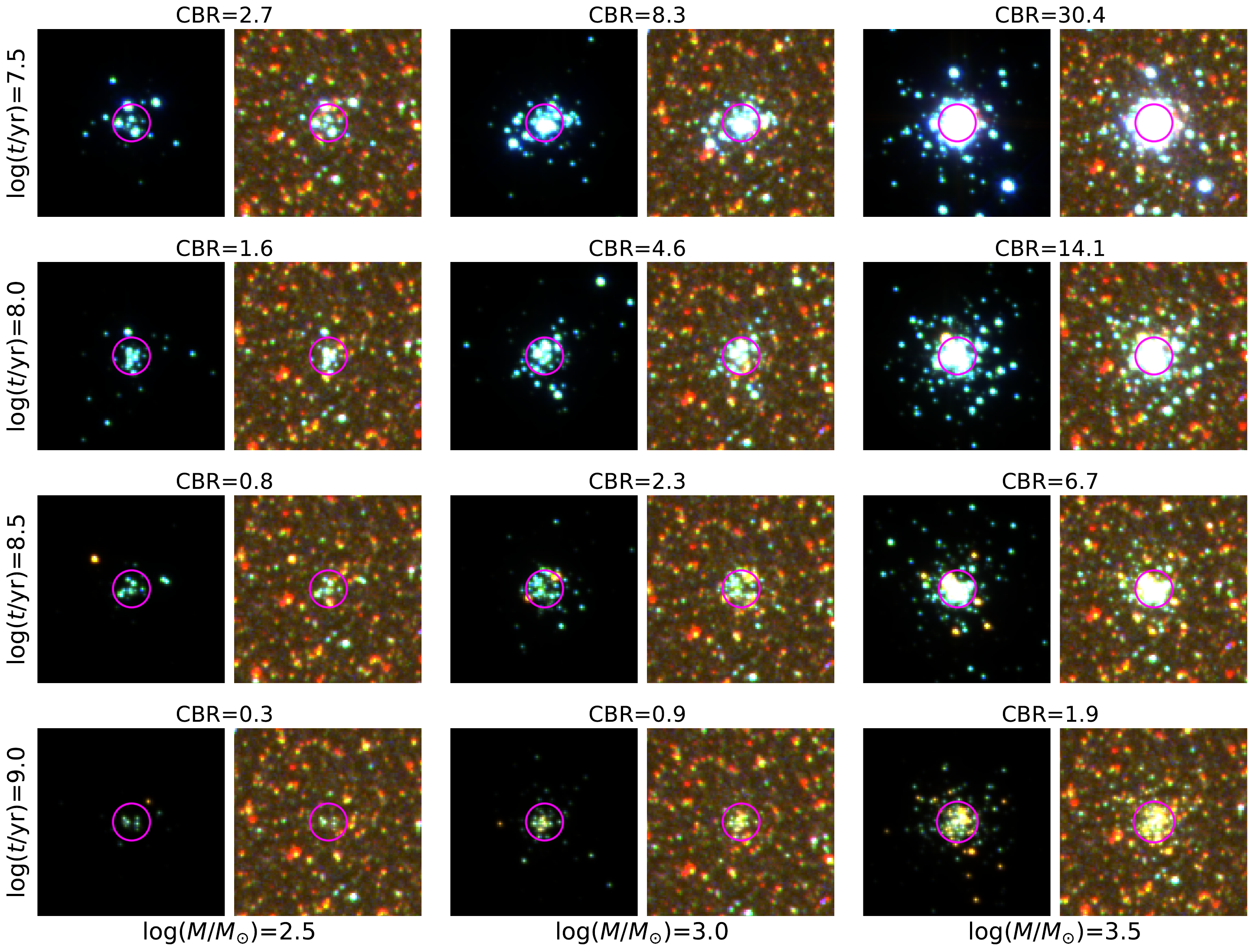}
	\caption{Images of artificial star clusters of various ages and masses in the B0 (left) and B3 (right) fields. Cluster mass is indicated below each column, and age is noted to the left of each row. All clusters were generated using $r_{c}=0.2$\arcsec\ and $\gamma=2.8$ parameters. The corresponding CBR values (see definition in Sect.\,\ref{Sec52}) determined in the B3 field are indicated above the image pairs. Magenta circles mark apertures used to measure CBR values. The colour images ($5\arcsec \times 5\arcsec$) were produced by combining $F336W$, $F475W$, and $F814W$ data.}
	\label{fig7}
\end{figure*}

\begin{figure}
	\centering
	\includegraphics[width=8.5cm]{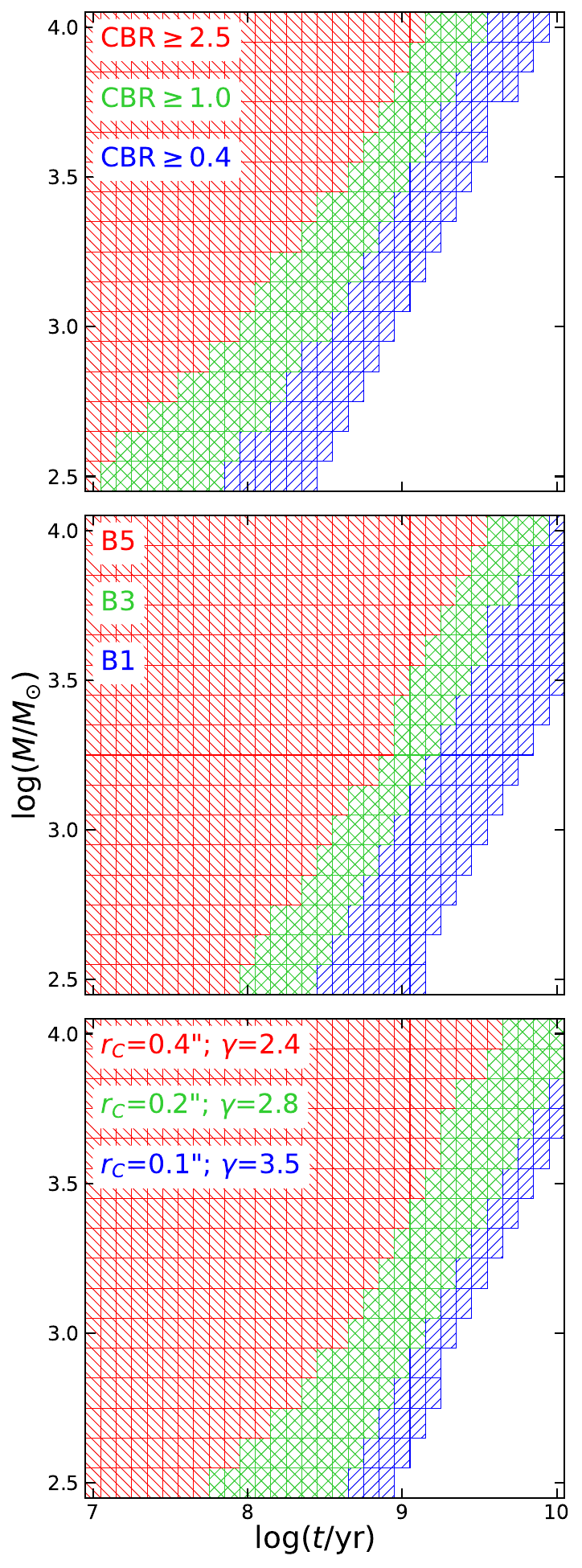}
	\caption{Cluster age-mass parameter space where at least 50\% completeness can be achieved. Top panel: dependence on CBR, with results shown for the B3 field. Middle panel: completeness limits across three background fields, assuming the detection threshold of $\rm CBR\geq0.4$. Bottom panel: results for three different cluster geometric parameter combinations in the B3 field, also assuming $\rm CBR\geq0.4$. Results shown in top and middle panels were determined using artificial clusters with $R_{50}= 0.25\arcsec - 1.00$\arcsec.}
	\label{fig8}
\end{figure}

\begin{figure}
	\centering
	\includegraphics[width=9.0cm]{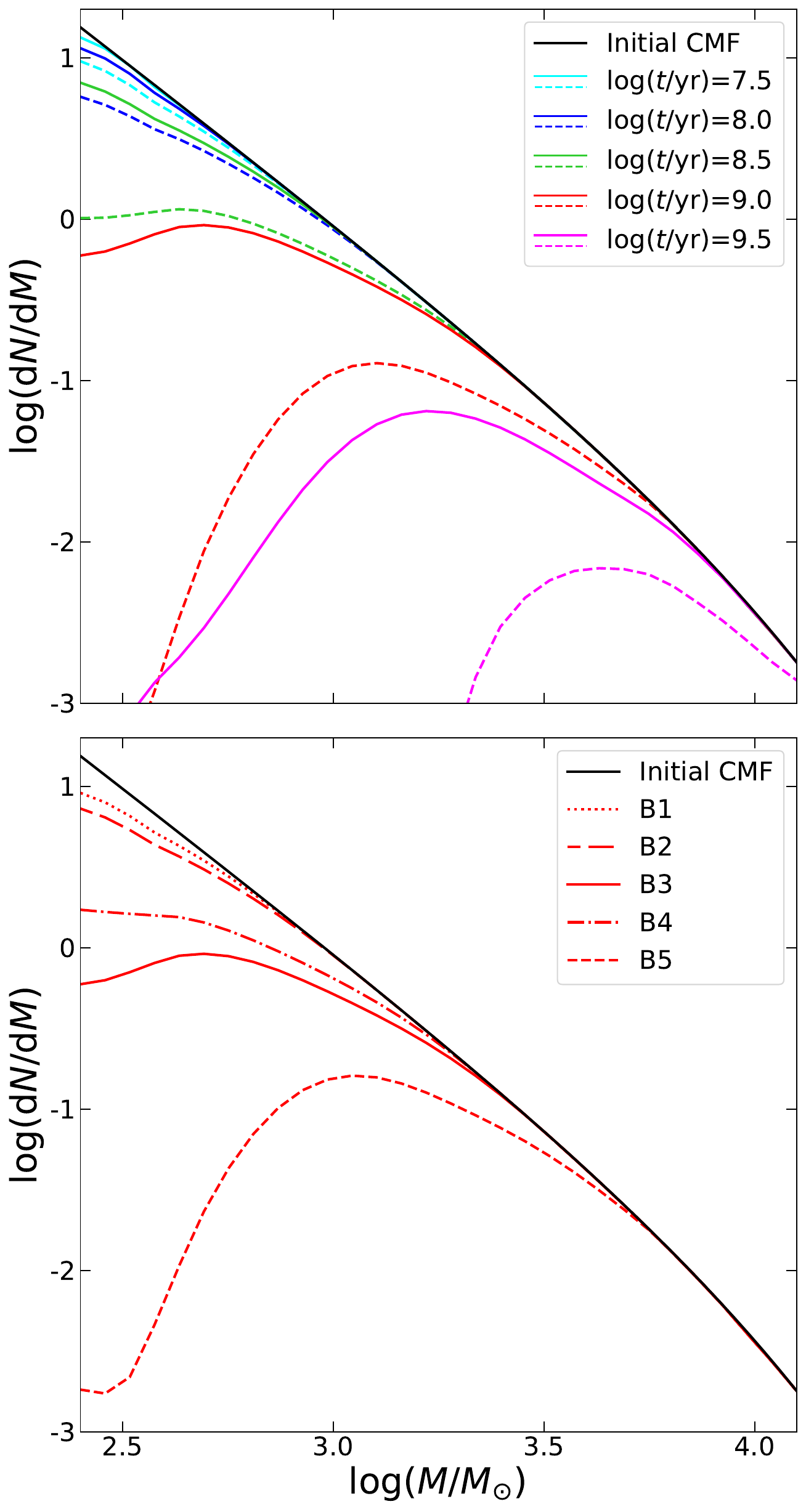}
	\caption{Selection effects on CMFs due to detection incompleteness. The initial CMF is shown as a black line (Schechter function with power-law index $\beta=-2$ and truncation mass $M_{\rm C}=10^4$~${M_\sun}$). Top panel: CMFs for artificial clusters of various ages (see legend) in the B3 field, showing results for $\rm CBR\geq1.0$ (dashed lines) and $\rm CBR\geq0.4$ (solid lines). Bottom panel: CMFs for clusters of log$(t/\rm yr)=9$ age, measured in the B1--B5 background fields at different radial distances from the \object{M\,31} centre, assuming the detection threshold $\rm CBR\geq0.4$. Results are based on artificial clusters with half-light radii $R_{50}=0.25\arcsec - 1.00$\arcsec.}
	\label{fig9}
\end{figure}

In this section, we explore two key aspects of background influence: its impact on completeness limits and on cluster detection quality. These factors define the accessible parameter space for star cluster studies and must be understood to avoid over-interpreting survey data.

To determine cluster detection completeness limits in the PHAT survey, the cluster-to-background signal ratio (CBR) was defined as:
\begin{fleqn}
	\begin{align}
	\label{eq:7} \mathrm{CBR} = \frac{{F_{cl}}}{F_{bg}}
	\end{align}
\end{fleqn}
where, $F_{cl}$ is the cluster's $F475W$ flux measured in the background-less field (B0) using an aperture $R_{\rm ap}=R_{50}$, and $F_{bg}$ -- the background $F475W$ flux measured within the same aperture in the selected \object{M\,31} fields (see Fig.\,\ref{fig1}). For CBR evaluation, we used the half-light radius aperture $R_{50}$, as it captures only the brightest central region of the cluster, providing the strongest contrast against the background, which is most significant for cluster detection. In cases where $R_{50}<0.5$\arcsec, we set $R_{\rm ap}=0.5$\arcsec\ to avoid the extreme stochastic effects associated with small apertures, as discussed in Sect.\,\ref{Sec4}. 

In Fig.\,\ref{fig7}, we show artificial star cluster images of various age and mass combinations placed in the B0 and B3 fields. For each cluster in the B3 field, the corresponding CBR value is indicated.

Further, we use a set of CBR threshold values as cluster detection criteria: $\rm CBR\geq2.5$, $\rm CBR\geq1.0$, and $\rm CBR\geq0.4$. As shown in Fig.\,\ref{fig7}, clusters with $\rm CBR\geq1.0$ can be visually detected with relative ease. We therefore adopt this threshold as a detection limit achievable through visual searches, such as those used in citizen science projects \citep{Johnson2015}. The threshold $\rm CBR\geq2.5$ corresponds to clusters that are $\sim$1~mag brighter, and $\rm CBR\geq0.4$ to those that are $\sim$1~mag fainter. We assume that the threshold $\rm CBR\geq0.4$ represents the PHAT survey depth achievable with current machine-learning (ML) algorithms. \citet{Thilker2022} showed that ML-produced star cluster catalogues in nearby galaxies can reach depths $\sim$1~mag fainter than those achieved through human detection.

In Fig.\,\ref{fig8}, we show the regions in the cluster age-mass parameter space where at least 50\% completeness can be achieved (in the context of the PHAT survey), depending on the selected CBR threshold, the background environment, and the cluster’s geometric parameters.

In the top panel of Fig.\,\ref{fig8}, we show results for the three aforementioned detection thresholds in the B3 field (clusters with $R_{50}=0.25\arcsec - 1.00\arcsec$\ are used). For the detection threshold $\rm CBR\geq2.5$ -- corresponding to clusters that can be studied with high accuracy -- the 50\% completeness limits are reached at $\sim$1300~${M_\sun}$, $\sim$2500~${M_\sun}$, and $\sim$8000~${M_\sun}$ (for ages 100~Myr, 300~Myr, and 1~Gyr, respectively). At a lower threshold of $\rm CBR\geq1.0$ -- corresponding to clusters detectable through visual searches -- these limits shift to $\sim$500~${M_\sun}$, $\sim$1000~${M_\sun}$, and $\sim$3000~${M_\sun}$ for the same ages. If the detection algorithm is capable of reaching $\rm CBR\geq0.4$, the resulting cluster catalogue becomes significantly deeper, with 50\% completeness achieved at $\sim$250~${M_\sun}$, $\sim$400~${M_\sun}$, and $\sim$1300~${M_\sun}$ (for ages 100~Myr, 300~Myr, and 1~Gyr, respectively). This represents an improvement of more than two times in the detectable mass range compared to the $\rm CBR\geq1.0$ cases. Interestingly, \citet{Thilker2022}, in the context of the PHANGS-HST survey, reported a similar gain when comparing the performance of their advanced ML-based cluster detection algorithm with expert scientists' visual identification.

The middle panel of Fig.\,\ref{fig8} shows variations of the 50\% completeness limits depending on the environment (radial distance from the \object{M\,31} centre). Here, results are presented assuming the detection threshold of $\rm CBR\geq0.4$ and using clusters of typical size ($R_{50}=0.25\arcsec - 1.00$\arcsec). Across the full range of environments -- from the sparse outer regions to the dense inner disc fields -- we find that the achievable survey depth at a certain age generally changes by $\sim$1~dex in mass. For example, at 1~Gyr, the 50\% completeness limit is $\sim$250~${M_\sun}$ in the B1 field, $\sim$1300~${M_\sun}$ in the B3 field, and $\sim$3000~${M_\sun}$ in the B5 field. These findings highlight the importance of carefully accounting for selection effects when investigating radial or environmental trends, as the completeness of a cluster catalogue strongly depends on the local background.

The bottom panel of Fig.\,\ref{fig8} illustrates the dependence of completeness limits on the geometric parameters of clusters, showing results for three $r_{c}$ and $\gamma$ combinations in the B3 field, assuming a detection threshold of $\rm CBR\geq0.4$. A significant selection bias toward more compact clusters is evident: for extended clusters (red area), the region of parameter space where at least 50\% completeness is achieved is substantially shallower compared to typical \object{M\,31} (green) and compact (blue) clusters. For example, at an age of 1~Gyr, the 50\% completeness limit is $\sim$400~${M_\sun}$ for compact clusters, $\sim$800~${M_\sun}$ for average ones, and only $\sim$2500~${M_\sun}$ for extended clusters.

This highlights the importance of carefully accounting for selection effects when studying phenomena related to cluster size -- such as the cluster size distribution, mass-radius relation, dynamical evolution, and tidal disruption \citep{PortegiesZwart2010, Krumholz2019, Adamo2020, Brown2021}. In (semi-)resolved cases, the completeness of star cluster catalogues can be biased towards more compact objects, particularly in nearby galaxies like \object{M\,31} or \object{M\,33}. On the other hand, in more distant galaxies, catalogue completeness could be biased towards larger clusters, as compact clusters are often unresolved and can be misclassified as point sources due to limited spatial resolution \citep{Ryon2017, Brown2021, Maschmann2024}.

\citet{Johnson2015} reported that the PHAT cluster catalogue is 50\% complete down to $\sim$500~${M_\sun}$ for ages younger than 100~Myr. Our results, for 100~Myr old clusters and a $\rm CBR\geq1.0$ detection limit, are in general agreement, at $\sim$500~${M_\sun}$. Thus, our assumption that the $\rm CBR\geq1.0$ limit is similar to human-eye abilities appears to be reasonable. However, while \citet{Johnson2015} includes all clusters younger than 100~Myr, our estimate is for an exact artificial cluster age of 100~Myr. The assumed $\rm CBR\geq1.0$ detection limit might be slightly looser than the one achieved by citizen science in \citet{Johnson2015}. Therefore, it seems that the PHAT cluster catalogue could be expanded at the low-mass ($\lesssim$500~${M_\sun}$) end by employing advanced ML-based cluster detection methods \citep{Bialopetravicius2019, Thilker2022}. 

Furthermore, we use CBR measurements to estimate the impact of selection effects on the cluster mass function (CMF) in the context of the PHAT survey (Fig.\,\ref{fig9}). As the initial CMF, we adopt the Schechter function with a power-law index of $\beta=-2$ \citep{Schechter1976, Gieles2009} and truncation mass of $M_{\rm C}=10^4$~${M_\sun}$, consistent with results for \object{M\,31} \citep{Johnson2017}.

In the top panel of Fig.\,\ref{fig9}, we show CMFs for several cluster ages, each shown under two detection thresholds: $\rm CBR\geq1.0$ (dashed lines) and $\rm CBR\geq0.4$ (solid lines). In the case of the $\rm CBR\geq0.4$ threshold, for clusters younger than 100~Myr, the CMF can be recovered down to $\sim$300~${M_\sun}$, and for clusters of 300~Myr ages the reachable mass limit could be $\sim$600~${M_\sun}$. 

The bottom panel of Fig.\,\ref{fig9} shows the 1~Gyr age CMF dependence on the environment (radial distance from the \object{M\,31} centre), assuming a cluster detection threshold of $\rm CBR\geq0.4$. We find that the CMFs can be recovered down to mass of $\sim$500~${M_\sun}$ in the outer disc regions (the B1 and B2 fields ), and to $\sim$1500~${M_\sun}$ in the intermediate disc areas or star-forming regions (the B3 and B4 fields). In contrast, in the dense inner disc environment (the B5 field), the CMF is only complete down to $\sim$3000~${M_\sun}$. We note that the flat tail at $\sim$300~${M_\sun}$ observed in the B5 field is statistical in nature and arises from the limited number of low-mass clusters measured. 

It is important to note that interstellar extinction effects were not included in our modelling of completeness and selection effects. However, as discussed by \citet{Maschmann2024}, a significantly higher incompleteness can be expected in UV-optical-NIR observations of very young clusters ($\lesssim$5~Myr), as they often are enshrouded in their natal gas and dust clouds. To avoid presenting potentially misleading information, in Fig.\,\ref{fig8} and Fig.\,\ref{fig9} we do not provide results for clusters younger than 10~Myr.

\section{Summary and conclusions}
\label{Sec6}

In this study, we investigated the accuracy and applicability limits of the aperture photometry method for star clusters, focusing on the effects of background and stochasticity. Here, we expand on our previous study \citepalias{Daugevicius2024}, where we mainly explored the impact of stochasticity. We use a large grid of artificial star clusters covering the parameter space of real objects observed in \object{M\,31}. Cluster images were generated in six HST ACS+WFC3 passbands, mimicking star cluster observations from the PHAT survey \citep{Dalcanton2012, Johnson2012, Johnson2015}. To create realistic mock observations of clusters, we placed them into five background fields selected from the PHAT data, each representing a different stellar density and environment. We performed aperture photometry on the resulting images, estimated photometric uncertainties, determined optimal aperture sizes, and estimated deviations in photometric measurements introduced by background effects. Furthermore, using artificial  clusters, we investigated star cluster selection effects and completeness limits in the context of the PHAT survey.

For the PHAT survey, we identified regions of the age-mass parameter space of star clusters where photometric uncertainties are low enough to allow a simultaneous determination of age, mass, extinction, and metallicity \citep{deMeulenaer2014, deMeulenaer2015}. We also demonstrated the dependence of these regions on both the environment (radial distance from the \object{M\,31} centre) and the aperture size used for photometry. Additionally, we identified which cluster age and mass combinations in a typical \object{M\,31} environment assure a PHAT photometric accuracy sufficient for cluster parameter determination. We estimated that at 1~Gyr, photometry is accurate enough to determine the age, mass, and extinction (assuming known metallicity) for clusters $\gtrsim$1600~${M_\sun}$; however, only rough age and mass estimates are possible for clusters $500\lesssim {M} \lesssim 1600$~${M_\sun}$. At an age of 100~Myr, the corresponding mass limits are: $\gtrsim$500~${M_\sun}$ and $200 \lesssim {M} \lesssim 500$~${M_\sun}$.

Based on our estimates of photometric uncertainties, we derived optimal aperture radii $R_{\rm opt}$. We demonstrated that $R_{\rm opt}$ depends on cluster brightness, with more luminous objects requiring larger apertures. Within the cluster magnitude range explored in this paper, we show that $R_{\rm opt}$ varies by a factor of $\sim$2, despite identical cluster geometric parameters. For CI measurements, we find that the optimal aperture radius for typical-size clusters, $\sim$(1--2)$\times R_{50}$, is consistent with our recommendations from \citetalias{Daugevicius2024}. After evaluating background effects, we reiterate that using small apertures ($R_{\rm ap}\lesssim R_{50}$) should be avoided due to strong stochastic and background-induced uncertainties.

Moreover, we investigated the biases in cluster magnitude and CI measurements introduced by background effects for \object{M\,31} star clusters. We find that, beyond a certain cluster brightness threshold, systematic biases begin to appear -- clusters tend to be measured brighter and redder than they truly are. These deviations increase as cluster brightness decreases. Additionally, larger clusters are more susceptible to these effects. Such biases can lead to significant inaccuracies in the derived physical parameters of star clusters.

Furthermore, we investigated star cluster selection effects in the PHAT survey and identified the regions in the cluster age-mass parameter space where at least 50\% completeness limits can be achieved. For the PHAT survey, we derive that the 50\% completeness limits could be reached at $\sim$250~${M_\sun}$ for clusters of 100~Myr, $\sim$400~${M_\sun}$ at 300~Myr, and $\sim$1300~${M_\sun}$ at 1~Gyr.

We show that within the context of the PHAT survey, a significant selection bias towards more compact clusters may exist. Thus, when studying various relations connected with cluster size -- such as structural evolution, disruption, or scaling relations -- it is necessary to account for these selection effects.

Finally, we estimated the impact of selection effects on the CMF in the \object{M\,31} disc. We derived that the CMF at 1~Gyr can be recovered down to: $\sim$500~${M_\sun}$ in the outer regions of the disc; $\sim$1500~${M_\sun}$ in dense or star-forming regions; $\sim$3000~${M_\sun}$ in the inner disc regions. These results stress the need to carefully account for selection effects when studying radial or environmental dependencies of the CMF.

We anticipate that the results presented in this study will be valuable for planning future observations and for interpreting results from current and upcoming star cluster surveys in the local universe. In the future, we aim to apply the findings from \citetalias{Daugevicius2024} and this study to develop automated, ML-based cluster detection and aperture photometry algorithms.  Such tools will be especially important in the era of Big Data Astronomy, where large wide-field star cluster surveys will make manual, interactive approaches impractical.

\begin{acknowledgements}
This research has made use of: the NASA/IPAC Extragalactic Database (NED), which is funded by the National Aeronautics and Space Administration and operated by the California Institute of Technology; SAOImage DS9, developed by the Smithsonian Astrophysical Observatory (\url{https://ds9.si.edu}); Astropy (\url{http://www.astropy.org}), a community-developed core $\tt Python$ package for Astronomy \citep{Astropy2013, Astropy2018}; $\tt APLpy$, an open-source plotting package for $\tt Python$ \citep{aplpy2012}; SciPy, an open-source scientific library for $\tt Python$ \citep{scipy2020}; Matplotlib, a plotting library for $\tt Python$ \citep{Hunter2007}; TOPCAT (\url{https://www.star.bristol.ac.uk/mbt/topcat}). It is based on observations made with the NASA/ESA $Hubble$ Space Telescope, and obtained from the $Hubble$ Legacy Archive, which is a collaboration between the Space Telescope Science Institute (STScI/NASA), the Space Telescope European Coordinating Facility (ST-ECF/ESA) and the Canadian Astronomy Data Centre (CADC/NRC/CSA). This project has received financial support from the Research Council of Lithuania (LMTLT), agreement No S-MIP-24-98. Computations were performed on the supercomputer GALAX of the Center for Physical Sciences and Technology, Lithuania. 
\end{acknowledgements}

\bibliographystyle{aa}

\end{document}